\documentclass[showpacs,preprintnumbers,amsmath,amssymb,twocolumn,superscriptaddress,prb]{revtex4}
\usepackage{times}
\usepackage{graphicx}
\usepackage{amsfonts}
\usepackage{amsmath, amsthm, amssymb}
\usepackage{dsfont}
\usepackage{mathptm}
\usepackage{color}

\definecolor{Red}{rgb}{1,0,0}

\newcommand{\s}{\mathbf{s}} 

\begin{document}
\title[Controllable Spin-Transfer Torque on an Antiferromagnet in a Dual Spin-Valve]{Controllable Spin-Transfer Torque on an Antiferromagnet in a Dual Spin-Valve}

\author{Jacob Linder}
\affiliation{Department of Physics, Norwegian University of Science and Technology, N-7491 Trondheim, Norway}

\date{Received \today}
\begin{abstract}
We consider current-induced spin-transfer torque on an antiferromagnet in a dual spin-valve setup. It is demonstrated that a net magnetization may be induced in the AFM by partially or completely aligning the sublattice magnetizations via a current-induced spin-transfer torque. This effect occurs for current densities ranging below 10$^6$ A/cm$^2$. The direction of the induced magnetization in the AFM is shown to be efficiently controlled by means of the magnetic configuration of the spin-valve setup, with the anti-parallell configuration yielding the largest spin-transfer torque. Interestingly, the magnetization switching time-scale $\tau_\text{switch}$ itself has a strong, non-monotonic dependence on the spin-valve configuration. These results may point toward new ways to incorporate AFMs in spintronic devices in order to obtain novel types of functionality. 

\end{abstract}

\maketitle
\section{Introduction}
The effect known as spin-transfer torque \cite{slon, berger_prb_96} lies at the heart of many areas within theoretical and applied spintronics \cite{rmp}. In essence, it consists of non-equilibrium spin-polarized electrons transferring angular momentum onto a magnetic order parameter. For a sufficiently large current of such electrons, one may for instance observe magnetization switching of a magnetic layer. A flow of electrons can thus serve to manipulate the configuration of magnetic hybrid structures, which is a key element in modern spintronics. 

Whereas current-induced spin-transfer torque traditionally has been discussed in the context of ferromagnetic layers, there has arisen an interest in how this phenomenon is manifested in antiferromagnets (AFMs) \cite{nunez_prb_06, tang_apl_07, urazhdin_prl_07, wei_prl_07, zu_prl_08, herranz_prb_09, gomonay_prb_10, hals_prl_11, swaving_prb_11}. The motivation for this pertains to both fundamental and applied aspects. Regarding the former, the ambition is to extend the concept of spin-transfer torque to systems with different types of magnetic ordering than ferromagnets. Concerning the latter, it has been demonstrated that the current density required to influence the exchange bias of antiferromagnets can be 1-2 orders of magnitude smaller \cite{wei_prl_07, dai_prb_08} than for magnetoresistive structures consisting of ferromagnets (FMs). Such a property is highly desirable in the context of practical devices. Related to this, it is also known \cite{berger_jap_03} that the critical current may be decreased considerably by employing a so-called dual spin-valve setup where a free layer is sandwiched between two polarizing magnets that both can act with a spin-transfer torque on the middle layer.

Taking the above observations into account, an interesting opportunity presents itself: how does a current-induced spin-transfer torque act on an antiferromagnet in a dual-spin valve setup? Given the advantages for both AFMs and dual-spin valves mentioned previously, one might expect that a combination of these two aspects could generate new and improved perspectives in spintronics. Motivated by this, we consider in this Letter 
the magnetization dynamics induced by an electric current flowing through an antiferromagnet in a dual spin-valve setup. Our three main findings are the following: \textit{(i)} For an experimentally feasible parameter range, we find that it is possible to induce a net magnetization in the AFM by partially or completely aligning the sublattice magnetizations via a current-induced spin-transfer torque. \textit{(ii)} The direction of the induced magnetization in the AFM can be efficiently controlled by means of the magnetic configuration of the spin-valve setup, with the anti-parallell configuration yielding the largest spin-transfer torque. \textit{(iii)} The magnetization switching time $\tau_\text{switch}$ has a strong, non-monotonic dependence on the spin-valve configuration, suggesting that the switching time-scale itself can be tuned by varying the magnetization orientation of the spin-valve. These results may suggest novel routes to incorporating AFMs in spintronic devices with new types of functionality. 

\begin{figure}[t!] \centering \resizebox{0.48\textwidth}{!}{
\includegraphics{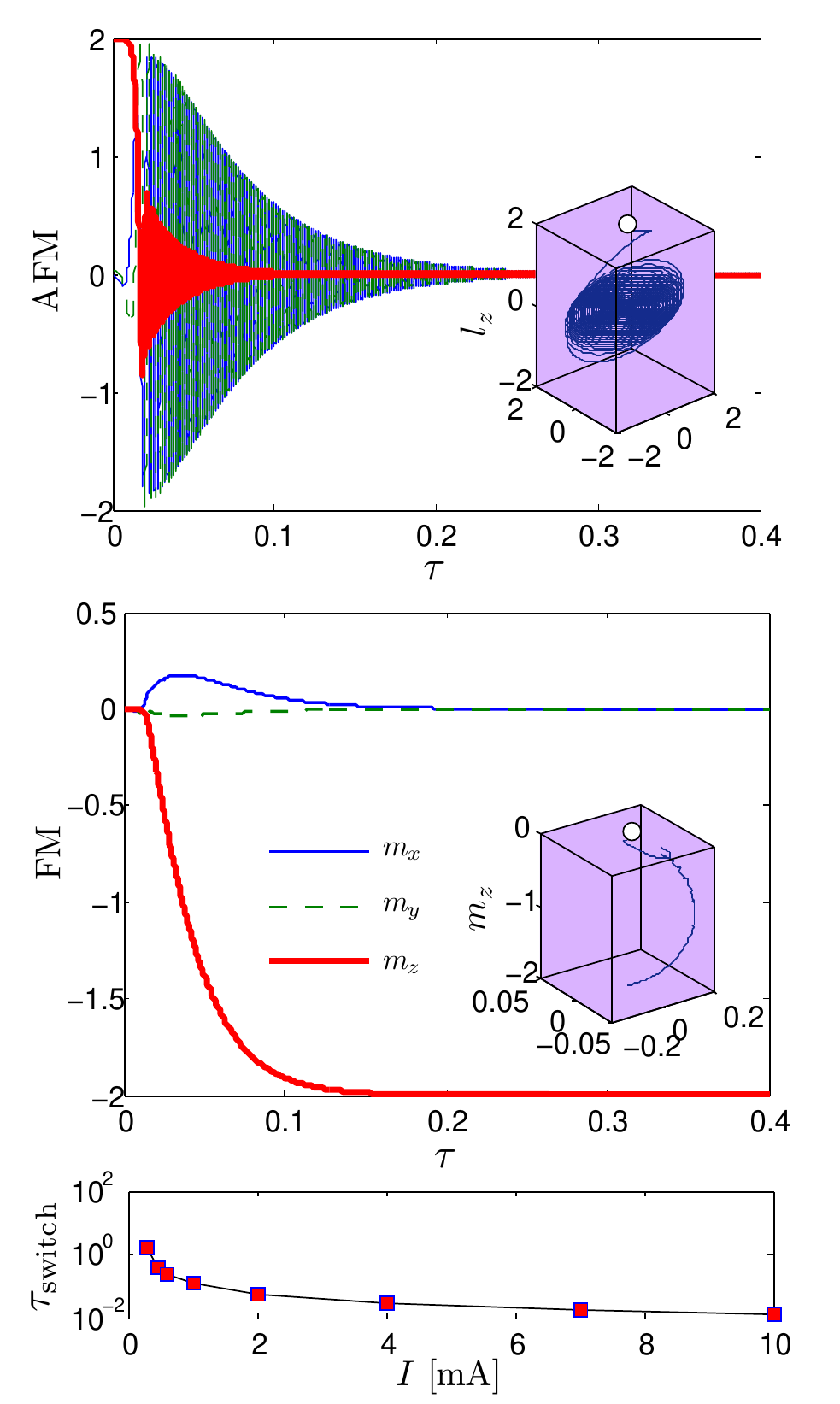}}
\caption{(Color online). Time-evolution of the AFM and FM order parameters $\mathbf{l}$ and $\mathbf{m}$, respectively, under the influence of an applied current-bias of $I=1$ mA. The spin-valve configuration is assumed to be AP ($\Omega=\pi$). Top panel: $\mathbf{l}$. Middle panel: $\mathbf{m}$. In both insets, a parametric plot is given with the circle indicating the order parameter value at $t=0$. Lower panel: switching time $\tau_\text{switch}$ and its dependence on the applied current-bias. 
} \label{fig:main}
\end{figure}

\section{Theory}
We now proceed to present the theoretical framework used to obtain these results. To study the time-dependent magnetization dynamics in the presence of a current-induced torque and anisotropy forces, we utilize the Landau-Lifshitz-Gilbert (LLG) equation \cite{llg} with two coupled magnetic sublattices. These describe the AFM order and are exchange coupled with an internal field $H_E$. The LLG-equation takes the form \cite{gomonay_prb_10, balaz_prb_09}:
\begin{align}\label{eq:llg}
\partial_t \mathbf{s}_j &= \alpha_j \mathbf{s}_j \times \partial_t\mathbf{s}_j -\gamma\mathbf{s}_j\times\mathbf{H}_\text{eff,j} + \boldsymbol{T}_j,
\end{align}
where the current-induced spin-transfer torque reads:
\begin{align}\label{eq:torque}
\boldsymbol{T}_j = I\zeta_j \Big(\s_j\times[\s_j\times(\s_L-\epsilon\s_R)]\Big).
\end{align}
Here, $\epsilon$ is an asymmetry factor accounting for any difference in polarization efficiency for the left and right ferromagnetic layers. The normalized magnetization vectors in the left and right parts of the spin-valve are 
\begin{align}
\s_L = (0,0,1),\; \s_R = (0,\sin\Omega,\cos\Omega)
\end{align}
such that the configuration is parallell (P) for $\Omega=0$ and anti-parallell (AP) for $\Omega=\pi$, while \begin{align}
\zeta_j=\frac{\nu\hbar\gamma}{2S_{0,j}Ve}.
\end{align}
Here, $e$ is the electron unit charge, $\nu$ is the polarization efficiency, $\hbar$ is Planck's constant, $\mu_0$ is the magnetic permeability, $S_{0,j}$ is the magnetization amplitude of sublattice $j$, $I$ is the applied current bias, and $V$ is the volume of the system.
The effective field $\mathbf{H}_\text{eff,j}$ acting on magnetic sublattice $j$ may be defined as: 
\begin{align}
\mathbf{H}_\text{eff,j} = -\partial \mathcal{F}/\partial \mathbf{S}_j
\end{align}
where $\mathcal{F}$ is the free energy per unit volume. Here, $\mathbf{S}_j = S_j\mathbf{s_j}$ and we assume $S_1\simeq S_2=S_0$. The free energy of the AFM is taken in the form \cite{gomonay_prb_10}:
\begin{align}
\mathcal{F} &= \frac{H_E}{4S_0}\mathbf{M}^2 + \frac{H_\text{an}^\perp}{S_0}(\mathbf{L})_x^2 - \frac{H_\text{an}^\parallel}{8S_0^3}[(\mathbf{L})_x^4+(\mathbf{L})_y^4+(\mathbf{L})_z^4] \notag\\
&- \mathbf{H}_0\cdot\mathbf{M},
\end{align}
where we have defined the FM and AFM order parameters:
\begin{align}
\mathbf{M} = \mathbf{S}_1+\mathbf{S_2},\; \mathbf{L} = \mathbf{S}_1-\mathbf{S_2}.
\end{align}
This corresponds to a tetragonal anisotropy with the easy axes ($y$ and $z$) in the AFM plane, and also incorporates the strong exchange coupling $H_E$ between the magnetic sublattices. The above constitutes a system of non-linear coupled equations for the magnetization $\mathbf{s}_j$ of sublattice $j$. In order to make contact with a realistic experimental situation, we now discuss the choices for parameter values. We set $2\mu_0S_0 = 0.1$ T, $\mu_0 = 4\pi\times10^{-7}$ Tm/A, $V = 120\times60\times1.5$ nm$^3$, and $\nu=0.3$ as a moderate estimate \cite{gomonay_07}. To model a realistic antiferromagnet, the spin exchange coupling and anisotropy fields are taken as $H_E=400$ T and $H_\text{an}^\perp = 0.01$ T, $H_\text{an}^\parallel = 0.02$ T \cite{kimel_nphys_09}. The Gilbert damping constant is set to $\alpha=0.001$ $(\alpha_1=\alpha_2$) and the sublattice magnetizations are assumed to be slightly shifted from their equilibrium position (AP to each other, $\s_1=-\s_2$) at $t=0$ by an angle $\theta/\pi=0.005$, and set $\epsilon=1$. We also introduce the spin-flop transition field 
\begin{align}
H_\text{sf} = 2\sqrt{H_\text{an}^\parallel H_\text{E}}
\end{align}
for later use, which is $\simeq$ 5 T with the above choice of parameters. The model under consideration is summarized in Fig. \ref{fig:model}. For details concerning the solution method of the LLG-equation, see the Appendix.

\begin{figure}[t!] \centering \resizebox{0.42\textwidth}{!}{
\includegraphics{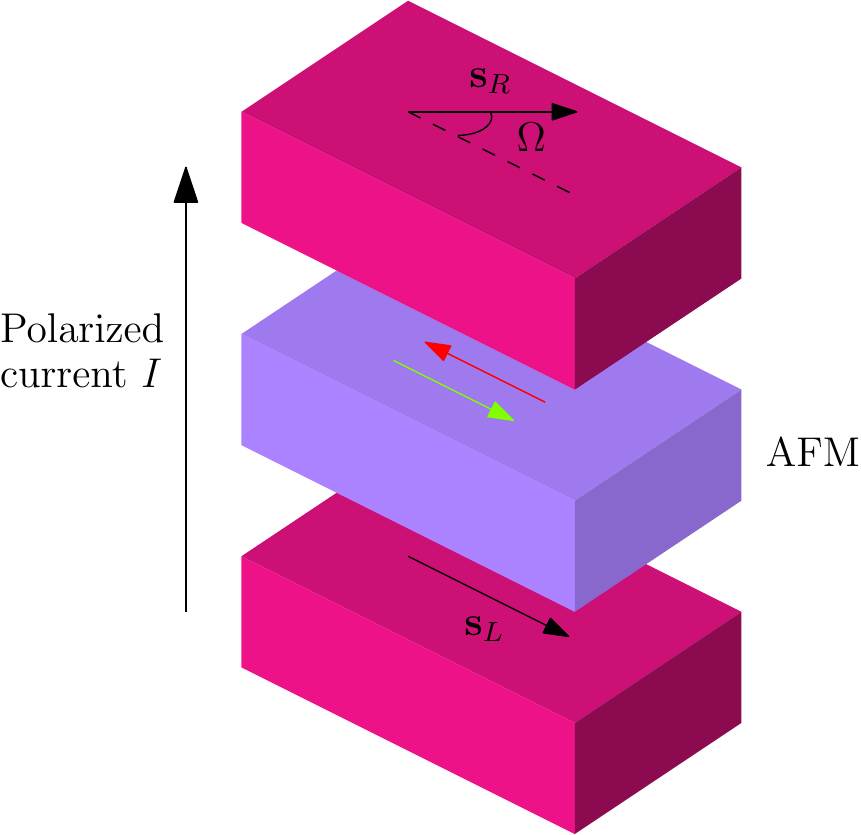}}
\caption{(Color online) Proposed setup of an antiferromagnetic layer sandwiched in a dual spin-valve setup. The magnetization orientations of the ferromagnetic layers may be misaligned with an angle $\Omega$. An injected electric current $I$ becomes polarized and transfers a spin-torque onto the magnetic sublattices in the AFM region, causing magnetization dynamics. The original equilibrium state of the AFM is indicated by the green and red arrows.
} \label{fig:model}
\end{figure}

\begin{figure}[t!] \centering \resizebox{0.48\textwidth}{!}{
\includegraphics{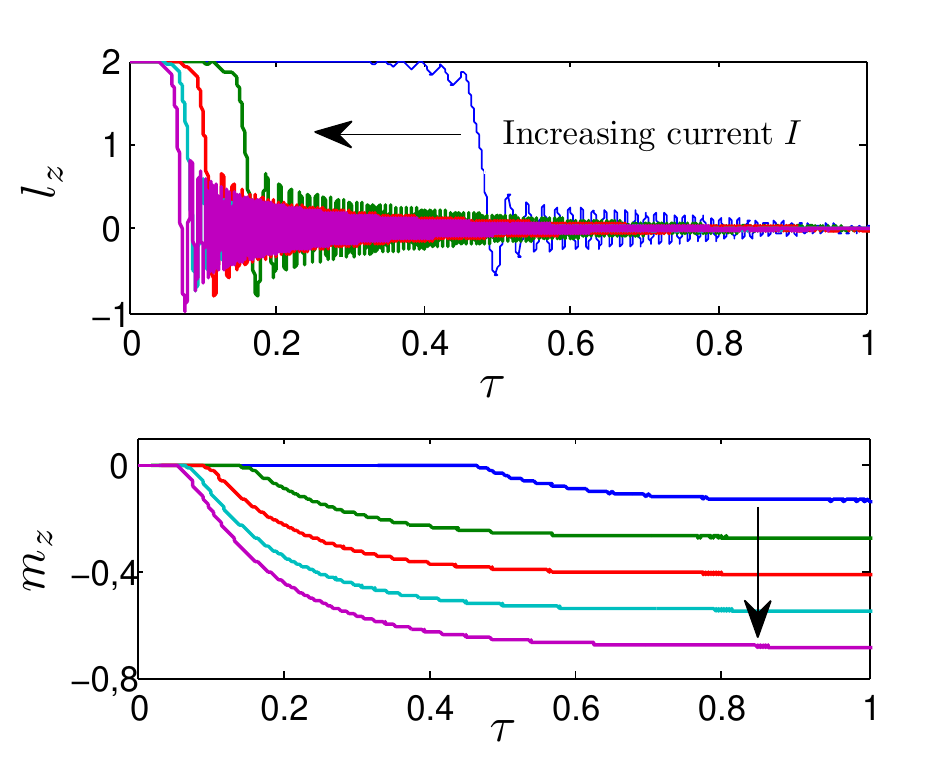}}
\caption{(Color online). Time-evolution for $l_z$ (left panel) and $m_z$ (right panel) in an AP spin-valve configuration ($\Omega=\pi$). The arrows point in the direction of increasing current: $I=\{0.02, 0.04, 0.06, 0.08, 0.10\}$ mA.
} \label{fig:smallcurrent}
\end{figure}

\section{Results}
In order to investigate quantitatively the magnetization dynamics, we have solved the full LLG-equation numerically. The time-coordinate has been normalized to $\tau = t\gamma\mu_0 M_0$ \cite{yan_arxiv_11}. We begin by focusing on the results obtained when the spin-valve configuration is AP, i.e. $\Omega=\pi$. The corresponding results are shown in Fig. \ref{fig:main} for a current bias of $I=1$ mA without any external field. As seen, the AFM order parameter $\mathbf{l}$ and FM order parameter $\mathbf{m}$ display qualitatively different behavior. In the top panel, $\mathbf{l}$ exhibits an oscillating decay until it vanishes completely. Remarkably, it is seen from the middle panel that a net magnetization evolves with increasing $\tau$ until it fully saturates in the $z$-direction. The insets show a parametric plot of the time-evolution of the AFM and FM order parameters, the circle denoting its value at $t=0$. These results indicate that it should be possible to magnetize an AFM exclusively by means of a current-bias in a spin-valve setup. In the lower panel of Fig. \ref{fig:main}, we consider how the switching time $\tau_\text{switch}$ depends on the magnitude of the applied current, the switching time defined from $|\mathbf{m}(\tau_\text{switch})| \geq 1.95$ (note that the maximum value of both $|\mathbf{m}|$ and $|\mathbf{l}|$ is 2). 

We proceed to investigate how large the current-bias has to be in order for the spin-transfer torque to magnetize the AFM. To answer this, we provide in Fig. \ref{fig:smallcurrent} both $l_z$ and $m_z$ as a function of $\tau$ for several values of the current strength $I$. As seen, the induced magnetization decreases as the current diminishes. However, even at $I=0.02$ mA one may observe a partial alignment of the sublattice magnetizations manifested as a finite value of $m_z$. As discussed in Sec. \ref{sec:discussion}, \textit{this corresponds to current densities ranging below 10$^6$ A/cm$^2$.} The AFM order parameter displays an oscillating decay in all cases. For very small currents $I\leq0.01$ mA, we found no appreciable induced magnetism when solving the above equations of motion. Instead, the AFM order parameter $\mathbf{l}$ undergoes a precessional motion and spin-flop transition into the $xy$-plane, similarly to Ref. \cite{gomonay_prb_10}.

\begin{figure}[t!] \centering \resizebox{0.48\textwidth}{!}{
\includegraphics{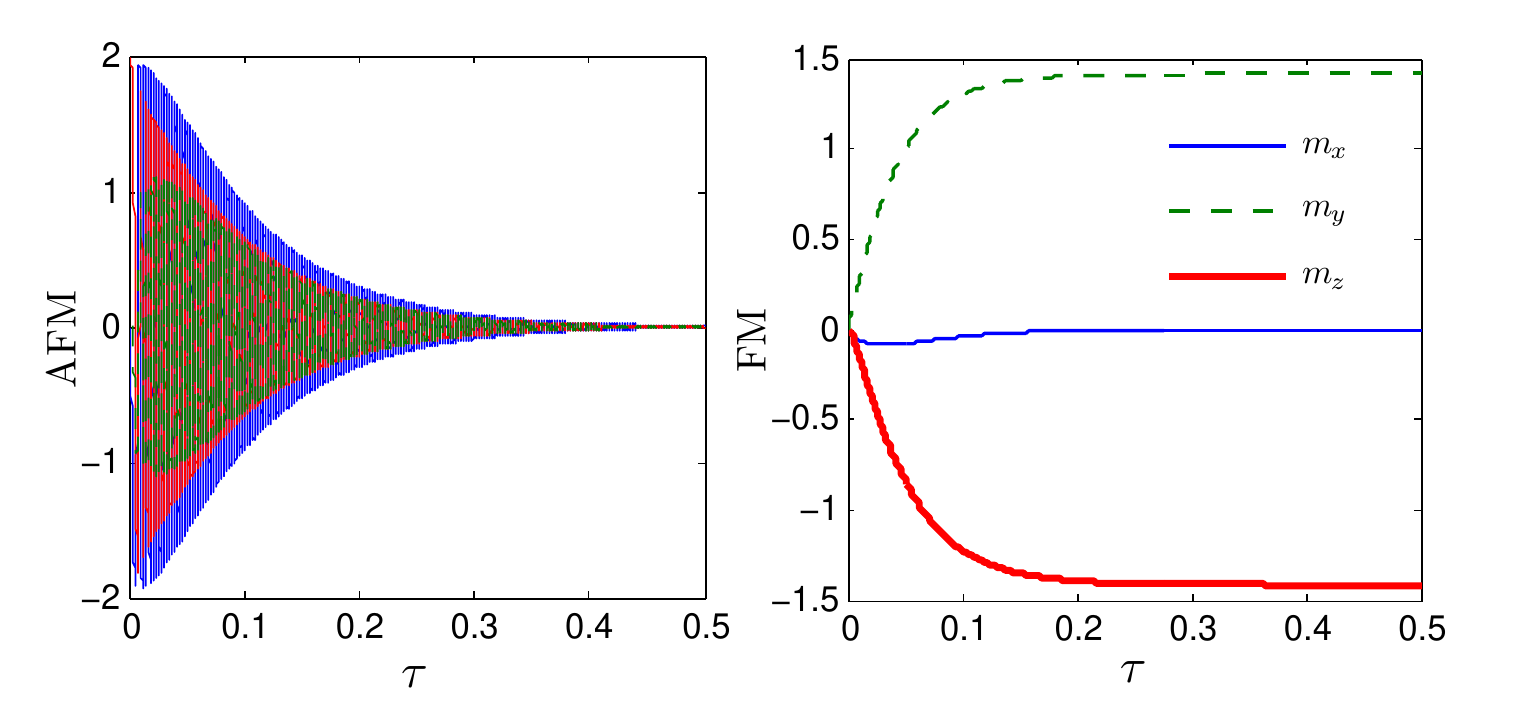}}
\caption{(Color online). Time-evolution of the AFM and FM order parameters $\mathbf{l}$ and $\mathbf{m}$, respectively, under the influence of an applied current-bias of $I=1$ mA. The spin-valve configuration is assumed to be set in an angle $\Omega=0.5\pi$. Left panel: $\mathbf{l}$. Right  panel: $\mathbf{m}$. 
} \label{fig:omega}
\end{figure}

It is also of interest to see what happens when the spin-valve configuration is noncollinear, i.e. $\Omega\neq \{0,\pi\}$. In Fig. \ref{fig:omega}, we solve for the time-evolution of the AFM and FM order parameters for an applied current of $I=1$ mA with a spin-valve configuration set at $\Omega=0.5\pi$. As seen, $\mathbf{l}$ decays to zero whereas $\mathbf{m}$ saturates at a finite value, albeit not fully aligned with either of the easy axes of the system. This shows that it is possible to control the direction of the induced magnetization in the AFM by tuning the spin-valve configuration $\Omega$. 

\begin{figure}[t!] \centering \resizebox{0.48\textwidth}{!}{
\includegraphics{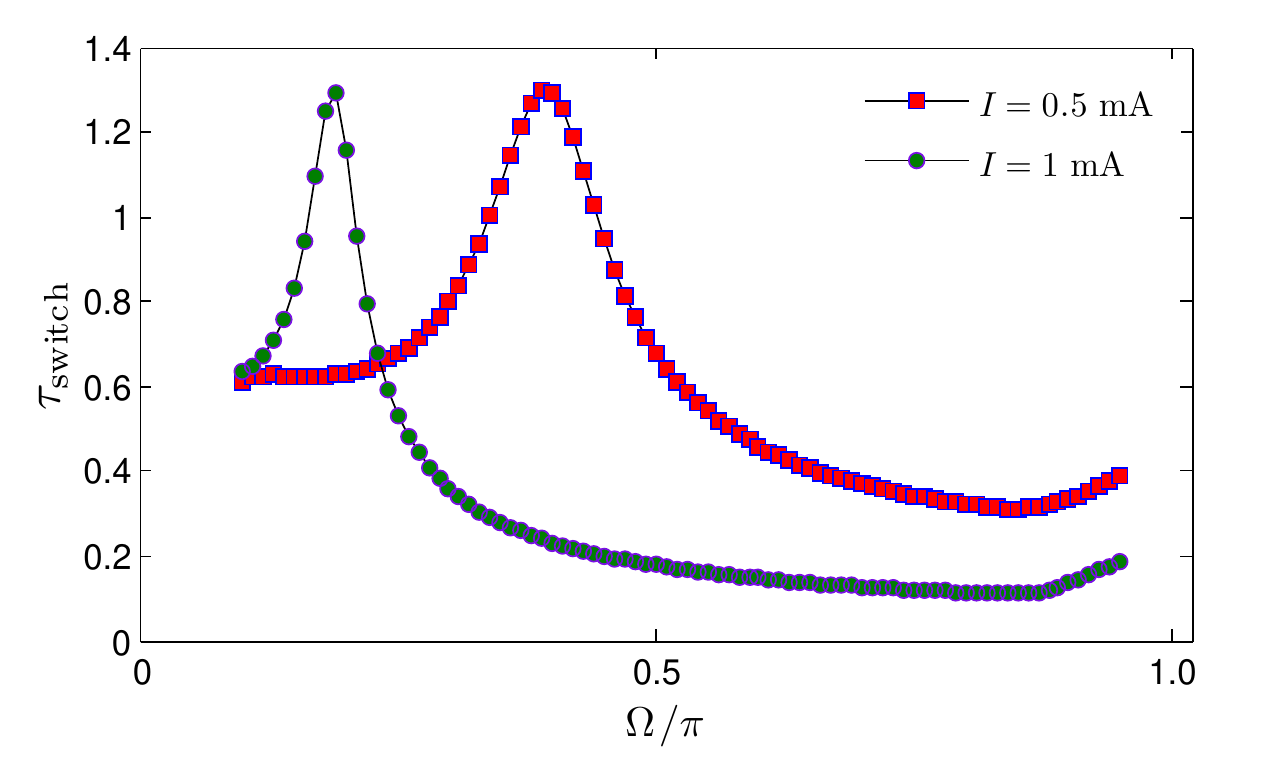}}
\caption{(Color online). Switching time and its dependence on the magnetic configuration of the spin-valve for a current bias of 0.5 mA and 1 mA, respectively. 
} \label{fig:switch}
\end{figure}

Related to this, it is natural to ask: what influence, if any, does the spin-valve configuration have on the switching time itself? To investigate this, we have plotted the switching time $\tau_\text{switch}$ as a function of $\Omega$ in Fig. \ref{fig:switch} for a regime of configurations where switching occurs, comparing two values of the current bias in our numerical calculations. Numerically, $\tau_\text{switch}$ was defined as the time where the magnetization had attained a value of $97.5\%$ of its saturated value. As seen, the switching time is strongly dependent on the configuration $\Omega$. In fact, it behaves in a non-monotonic fashion with a peak value as its most striking feature. Fig. \ref{fig:switch} suggests that there exists a spin-valve configuration for a given current bias which strongly delays the magnetization switching, whereas configurations close to AP ($\Omega=\pi$) offers the most rapid switching. The peak position shifts towards the P  configuration ($\Omega=0$) with increasing current which also lowers the overall switching time, as is natural since the spin-transfer torque becomes stronger. The variation in switching time as obtained when varying $\Omega$ is seen to span over more than an order of magnitude from Fig. \ref{fig:switch}.

\section{Discussion}\label{sec:discussion}

In order to understand the induced magnetic moment in the AFM qualitatively, one should note that the current-induced spin-transfer torque described by Eq. (\ref{eq:torque}) acts in the same direction even after applying the transformation $\s_j \to -\s_j$. Hence, both magnetic sublattices in the AFM will experience a torque in the same direction upon application of a current-bias and thus inducing a net magnetic moment. The stability range of the induced moment, i.e. whether it persists over time, depends on the other system parameters such as anisotropy field and exchange bias, as we have discussed. To observe the proposed effects, it is necessary to experimentally adjust the spin-valve configuration $\Omega$. Presumably, this will be most efficiently done by selecting ferromagnets with different properties for the left and right layer. The coercive field should be weaker for the tunable (right) magnetic layer such that its orientation may be changed by application of a small external field. An exchange interaction with the fixed (left) magnetic layer should determine the initial orientation of the AFM order parameter $\mathbf{l}$, taken to be along the $z$-axis in this case, although this interaction should be sufficiently small that it may be disregarded under the influence of a current-bias. 

A key parameter in terms of the current-induced spin-transfer torque is the required current density to obtain the induced magnetization $\mathbf{M}$. As seen from Eqs. (\ref{eq:llg}) and (\ref{eq:torque}), the torque is proportional to both the current density and the cross-sectional area of the AFM while being inversely proportional to the total volume of the AFM. Assuming a cross-sectional area of 120$\times$60 nm$^2$, it follows that the predicted effects in this paper occur for current densities even below 10$^6$ A/cm$^2$ (corresponding to a total current $\simeq 0.1$ mA). To further characterize the robustness of the reported effects, such as the non-monotonous switching-time \footnote{We note that a non-monotonous torque-dependence on the magnetic configuration has been reported in ferromagnetic bilayers previously: A. Kovalev, G. Bauer, A. Brataas, Phys. Rev. B \textbf{73}, 054407 (2006); J. Xiao, A. Zangvill, M. D. Stiles, Phys. Rev. B \textbf{70}, 172405 (2004).}, it could be useful to a apply micromagnetic theory to the proposed spin-valve structure. Moreover, the characterization of other properties such as how the GMR is influenced by the antiferromagnetic layer could provide further insight in how the magnetic configuration interacts with the presence of AFM in the middle free layer.

\section{Summary}
In summary, we have calculated the magnetization dynamics of an antiferromagnet in a dual spin-valve setup, taking into account anisotropy effects and current-induced torques. We have shown that it is possible to induce a net magnetization in the AFM by partially or completely aligning the sublattice magnetizations via a current-induced spin-transfer torque. Moreover, the direction of the induced magnetization in the AFM can be efficiently controlled by means of the magnetic configuration of the spin-valve setup. Remarkably, the magnetization switching time-scale itself is found to be controllable via the spin-valve setup: it displays a highly non-monotonic dependence on the magnetization configuration. The obtained results appear in an experimentally feasible parameter regime, and may thus point toward new ways to incorporate AFMs in spintronic devices in order to obtain novel types of functionality.

\begin{center}
\bf{Appendix}
\end{center}

We here provide some additional details concerning the method of solution for the LLG-equation. By direct algebraic manipulation, one may write Eq. (\ref{eq:llg}) as:
\begin{align}
\hat{\mathcal{A}}\mathbf{x} = \mathbf{B}
\end{align}
where $\mathbf{x} = [\dot{s}_{1x}, \dot{s}_{1y}, \dot{s}_{1z}, \dot{s}_{2x}, \dot{s}_{2y}, \dot{s}_{2z}]^\mathcal{T}$ where $\mathcal{T}$ denotes the matrix transpose. We have defined the matrices:
\begin{align}
\hat{\mathcal{A}} = \begin{pmatrix}
\underline{\mathcal{A}}_1 & \underline{0} \\
\underline{0} & \underline{\mathcal{A}}_2 \\
\end{pmatrix},\; \underline{\mathcal{A}}_j = \begin{pmatrix}
-1 & -\alpha s_{jz} & \alpha s_{jy} \\
\alpha s_{jz} & -1 & -\alpha s_{jx} \\
-\alpha s_{jy} & \alpha s_{jx} & -1\\
\end{pmatrix},\; j=1,2.
\end{align}
in addition to $\mathbf{B} = [\mathbf{b}_1, \mathbf{b}_2]^\mathcal{T}$ with:
\begin{align}
\mathbf{b}_j = \begin{pmatrix}
\gamma(s_{jy}H_{jz} - s_{jz}H_{jy}) - T_{jx}\\
\gamma(s_{jz}H_{jx} - s_{jx}H_{jz}) - T_{jy}\\
\gamma(s_{jx}H_{jy} - s_{jy}H_{jx}) - T_{jz}\\
\end{pmatrix}=\begin{pmatrix}
b_{jx}\\
b_{jy}\\
b_{jz}\\
\end{pmatrix},\; j=1,2.
\end{align}
The above system of equations may then be solved to yield uncoupled equations in the time-derivative of the magnetization sublattices $\mathbf{s}_j$, which read:
\begin{align}
\dot{s}_{jx} = -\frac{1}{1+\alpha^2}&\Big[\alpha^2( s_{jx}s_{jz} b_{jz}+  s_{jx}^2b_{jx} +s_{jx}s_{jy} b_{jy}) \notag\\
&+ \alpha( s_{jy}b_{jz} -s_{jz}b_{jy}) + b_{jx}\Big]\notag\\
\dot{s}_{jy} = -\frac{1}{1+\alpha^2}&\Big[\alpha^2(s_{jy} s_{jz} b_{jz} + s_{jy}^2 b_{jy} + s_{jx}s_{jy} b_{jx}) \notag\\
&+ \alpha(s_{jz}b_{jx} -s_{jx}b_{jz}) + b_{jy} \Big] \notag\\
\dot{s}_{jz} = -\frac{1}{1+\alpha^2}&\Big[\alpha^2(  s_{jy}s_{jz}b_{jy} +s_{jz}^2 b_{jz} +s_{jx} s_{jz}b_{jx}) \notag\\
&+ \alpha(s_{jx}b_{jy} - s_{jy}b_{jx}) + b_{jz} \Big] \notag\\
\end{align}

\end{document}